\documentclass[%
 aip,
 amsmath,amssymb,
 reprint,%
]{revtex4-1}

\usepackage{graphicx}
\usepackage{dcolumn}
\usepackage{bm}

\usepackage{color}
\usepackage[normalem]{ulem}

\usepackage[utf8]{inputenc}
\usepackage[T1]{fontenc}
\usepackage{mathptmx}

\begin{document}

\preprint{AIP/123-QED}

\title[]{The Seven Deadly Sins:\\when computing crystal nucleation rates, the devil is in the details}
\author{Katarina E. Blow}
\affiliation{ 
Department of Physics, University of Warwick, Coventry, CV4 7AL, United Kingdom
}

\author{David Quigley}%
 \email{D.Quigley@warwick.ac.uk}
  \affiliation{ 
Department of Physics, University of Warwick, Coventry, CV4 7AL, United Kingdom
}
\author{Gabriele C. Sosso}
 \email{G.Sosso@warwick.ac.uk}
 \affiliation{ 
Department of Chemistry, University of Warwick, Coventry, CV4 7AL, United Kingdom
}%

\date{\today}

\begin{abstract}
The formation of crystals has proven to be one of the most challenging phase transformations to quantitatively model - let alone to actually understand - be it by means of the latest experimental technique or the full arsenal of enhanced sampling approaches at our disposal. One of the most crucial quantities involved with the crystallization process is the nucleation rate, a single, elusive number that is supposed to quantify the average probability for a nucleus of critical size to occur within a certain volume and time span. A substantial amount of effort has been devoted to attempt a connection between the crystal nucleation rates computed by means of atomistic simulations and their experimentally measured counterparts. Sadly, this endeavour almost invariably fails to some extent, with the venerable classical nucleation theory typically blamed as the main culprit. Here, we review some of the recent advances in the field, focusing on a number of perhaps more subtle details that are sometimes overlooked when computing nucleation rates. We believe it is important for the community to be aware of the full impact of aspects such as finite size effects and slow dynamics, that often introduce inconspicuous and yet non-negligible sources of uncertainty into our simulations. In fact, it is key to obtain robust and reproducible trends to be leveraged so as to shed new light on the kinetics of a process, that of crystal nucleation, which is involved into countless practical applications, from the formulation of pharmaceutical drugs to the manufacturing of nano-electronic devices.
\end{abstract}

\maketitle

\section{\label{sec:intro}Introduction}

Understanding crystal nucleation is one of the fundamental ambitions of physical chemistry~\cite{yoreo_principles_2003,yi_molecular_2012,karthika_review_2016,sosso_crystal_2016}. Far from being a theoretical curiosity, the kinetics of crystallization has a crucial impact on many natural phenomena, such as the formation of ice~\cite{bartels-rausch_chemistry:_2013} or the process of biomineralisation~\cite{weiner_crystallization_2011}, and on a diverse range of practical applications such as the formulation of pharmaceutical drugs~\cite{lee_crystal_2011} or the design of novel nanostructures for data storage~\cite{sosso_fast_2013}. 

If we were to ask the reader to pick a single observable to characterise the nucleation process, the so-called crystal nucleation rate $\mathcal{J}$ would probably turn out to be a very popular choice. $\mathcal{J}$ is a scalar quantity that represents the average probability, per unit time and per unit volume, for a critical crystalline nucleus to occur within the supercooled liquid or the supersaturated solution of interest. The apparently straightforward nature of $\mathcal{J}$ makes it ideal to compare the crystallization kinetics of the same system in different conditions or indeed between different systems as well. Crucially, $\mathcal{J}$ can be both measured experimentally and estimated by means of computer simulations, thus providing, in principle, a much sought after connection between reality and modelling. However, long-standing, major inconsistencies between experiments and simulations still persist as of today, despite the ever-growing capabilities of molecular simulations~\cite{karthika_review_2016,sosso_crystal_2016}. 

When dealing with estimates of crystal nucleation rates obtained by means of atomistic simulations, the community has identified several outstanding issues through the years. Amongst the usual suspects, we can find the intrinsic limitations of the models we use to describe the interactions between atoms, or particles, or molecules~\cite{zimmermann_nucleation_2015,haji-akbari_direct_2015,demichelis_simulation_2018}. In addition, the rather dated theoretical framework provided by classical nucleation theory (CNT), whilst proving to be remarkably accurate in many cases, has been repeatedly put into question within the last few years~\cite{smeets_classical_2017, vekilov_nonclassical_2020}.

Here, we are going to focus on some (seven) aspects of interest for molecular simulations aimed at computing the crystal nucleation rate. Some of these issues, such as the potentially slow dynamics of the system~\cite{kuhnhold_derivation_2019}, are often overlooked. Some others, for instance the existence of finite size effects~\cite{hussain_how_2021}, are very well known - and yet, the extent of their impact on the estimate of $\mathcal{J}$ is still largely unexplored. Understanding the contributing factors to the uncertainty characterising $\mathcal{J}$ is crucial, not only to keep working toward the perhaps overly ambitious goal of reconciling simulations with experiments, but to allow meaningful insight to be extracted from our simulations in the first place. Indeed, these are really exciting times for the community, as a number of recent works have provided an unprecedented level of detail into specific aspects (negatively) affecting the calculation of $\mathcal{J}$. From the necessity to account for the microscopic kinetics of the system to the need of thoroughly investigating the adequacy of the order parameters we use, we take stock of the state-of-the-art and offer an opinionated perspective as to potential future developments within the field.

The paper is organised as follows: after having set the boundaries of our discussion and briefly reviewed the computational tools out our disposal as well as the above mentioned usual suspects (Sec.~\ref{sec:comp}), we focus on seven different aspects that are detrimental to the calculation of $\mathcal{J}$ by means of enhanced sampling atomistic simulations (Sec.~\ref{sec:seven}). Finally, in Sec,~\ref{sec:future} we assess the challenges ahead and put forward several possibilities we hope the community will decide to explore in the near future. 

\section{\label{sec:comp}Computing nucleation rates}

\begin{figure}[htbp]
\begin{centering}
\centerline{\includegraphics[width=0.45\textwidth]{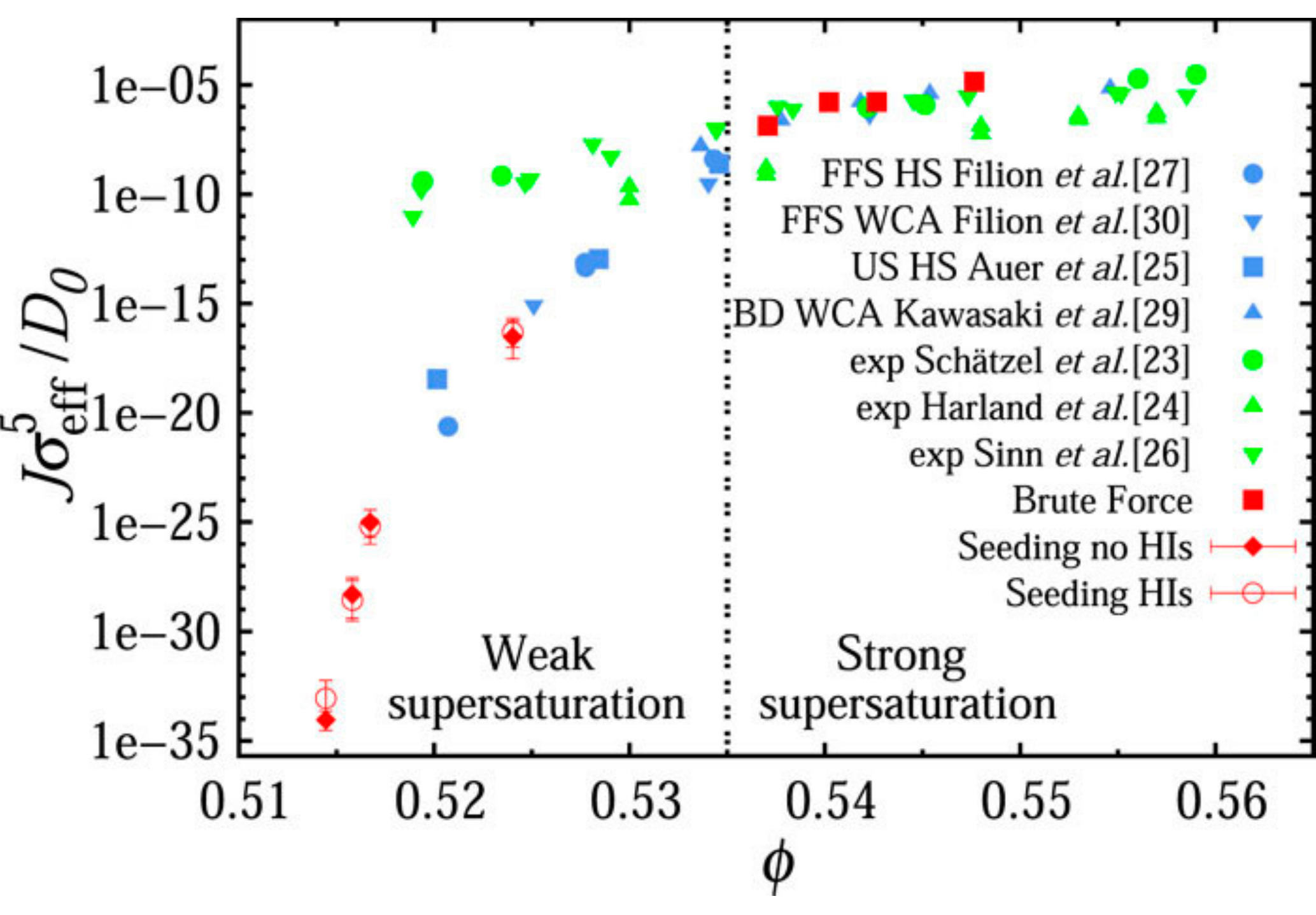}}
\par\end{centering}
\protect
\caption{\textit{The inconsistencies between experimental and simulated nucleation rates persists.} Nucleation rate $J\sigma^5_{\text{eff}}/D_0$ as a function of the volume fraction $\phi$ of a system of ``nearly'' hard spheres. This  is often considered as a ``model'' system, given the relatively straightforward interactions between the constituent particles. Experimental results are reported in green, while the outcomes of a variety of simulations are reported in blue or red. Note the substantial discrepancy ($\sim$10 orders of magnitude) between experiments and simulations in the weak supersaturation regime. The data points in red have been obtained in Ref.~\citenum{fiorucci_effect_2020} by means of three different approaches and are impressively consistent with each other, which is indicative of the fact that said discrepancy has nothing to do with the methodology employed. Adapted from Ref.~\citenum{fiorucci_effect_2020}}
\label{FIG_1}
\end{figure}

In order to discuss the fundamental aspects of calculating nucleation rates via atomistic simulations, we have to remove as many layers of complexity as possible from both the system under investigation and the conditions in which we are working on it. We start by focusing on supercooling or supersaturation regimes that lie well away from the spinodal limit where the free energy barrier associated with the formation of a critical nucleus is vanishingly small. Conversely, we have to avoid, by necessity, very mild supercooling or supersaturation as well; in that case, the size of the critical nucleus is usually too large to be even taken into account by means of atomistic simulations. Then, we shall consider almost exclusively homogeneous nucleation, in principle the simplest scenario available to us. However, homogeneous nucleation is rarely observed in reality, with heterogeneous nucleation being much more common. Indeed, it is often very challenging to  experimentally measure crystal nucleation rates without any influence in terms of impurities, which almost always manage to facilitate the crystallization process~\cite{karthika_review_2016,sosso_crystal_2016}. Note, however, that much of what is discussed in the following is equally relevant to studies of heterogeneous nucleation. In addition, we will largely avoid the emergence of two- or even multi-step nucleation processes (biomineralisation being a prominent example) as well as confinement effects. Excellent reviews on these topics and, broadly speaking, on the subject of crystal nucleation as a whole can be found in e.g. Refs.~\citenum{yi_molecular_2012,agarwal_nucleation_2014,karthika_review_2016} and~\citenum{vekilov_nonclassical_2020}. 

At this stage, it is useful to highlight the distinction between ``realistic'' and ``model'' systems. Simulating the former usually represents an attempt to get as close as possible to the experimental reality, by utilising the most accurate interatomic potentials available. Note that leveraging \textit{ab initio} simulations is, aside for very rare exceptions~\cite{sosso_fast_2013}, simply not feasible, as both the time- and length-scales involved in the crystal nucleation process (particularly taking into account its stochastic nature) lie far beyond the reach of these accurate, but rather computationally expensive approaches. Machine learning(ML)-based interatomic potentials~\cite{behler_perspective:_2016} represent an especially intriguing avenue to strike the balance between the computational efficiency of classical potentials and the accuracy of first principles calculations; however, we argue that, aside for exceptional cases where the kinetics of crystallization is unusually fast (see e.g. Ref.~\citenum{sosso_harnessing_2019}), these novel potential are still too computationally costly to be used to compute crystal nucleation rates with sufficient accuracy. Even when choosing the most accurate classical potentials at our disposal, the computational cost involved with the calculation of nucleation rates increases drastically, usually imposing a tough compromise between the quality of the simulations and the quantity of the results. In fact, the majority of rate calculations for realistic system (notable examples would be Refs.~\citenum{haji-akbari_direct_2015,sosso_microscopic_2016} and~\citenum{sosso_unravelling_2018}) focus on a single supercooling or supersaturation condition. 
Given that reaching an agreement between experimentally measured and simulated nucleation rates is, as we shall see shortly, very challenging, this is obviously sub-optimal. 

On the other hand, computing crystal nucleation rates for ``model'' systems aims at employing simple and/or computational inexpensive interatomic potentials, so as to explore different e.g. supercooling conditions and thus extract useful trends to be compared with the experimental data. Clearly, this often comes at the cost of sacrificing some accuracy in terms of the description of the actual system. However, it is important to point out that, in some cases, simple potentials can describe experimentally relevant systems to a very reasonable level of accuracy. Representative examples of such systems that we will often reference in this work are the Lennard-Jones (LJ) system and colloidal particles. These have all been extensively investigated through the years and thus offer the perfect opportunity to discuss several aspect related to the calculation of nucleation rates via atomistic simulations.

One issue that unfortunately unifies the vast majority of computational studies, not necessarily limited to those dealing with crystal nucleation rates, is the reproducibility of the results. This is especially evident in the case of LJ systems, where minimal modifications to details such as the truncation/shift (which vary wildly within the literature) lead to substantially different melting curves and nucleation rates. In fact, as we shall discuss in greater detail in Sec.~\ref{sec:force}, the estimate of $\mathcal{J}$ is incredibly susceptible to each and every feature of the interatomic potential of choice. In this context, it is comforting to witness the emergence of more and more open access databases where the actual input files used to obtain a particular computational results are reported in full (see e.g. Ref.~\citenum{bonomi_promoting_2019}), thus greatly facilitating cross-validation.

In contrast, one key issue that stubbornly refuses to yield to the efforts of the community is the the long-standing discrepancy between experimental and simulated crystal nucleation rates. Perhaps the starkest reminder of how much work remains to be done in the field is given by the current state of affairs with respect to colloidal systems. These qualify as ``simple'' systems, given that they can, in principle, be modelled  via very inexpensive models such as hard or soft spheres. Crucially, it is also possible to measure crystal nucleation rates of colloidal systems via relatively straightforward experimental techniques such as confocal microscopy~\cite{wang_direct_2015}. As can be seen in Fig.~\ref{FIG_1}, however, while experiments and simulations agree quite well at strong supersaturation, discrepancies as large as ten orders of magnitude can be found at weak supersaturation - thus indicating that, perhaps, our simple models cannot in fact be used to simulate these systems accurately enough. This evidence is quite astonishing, as we have by now reached a point where different approaches to the calculation of $\mathcal{J}$ are consistent with each other. In particular, the simulation results reported in Fig.~\ref{FIG_1} for weak supersaturation (obtained by Fiorucci \textit{et al.} in 2020~\cite{fiorucci_effect_2020}) have been validated by three different methodologies. The fact that the discrepancy between experiments and simulations still persists is thus suggestive of the possibility that there must be some fundamental aspect of our simulations that we are still missing - and that has nothing to do with the choice of the method used. Potential culprits are the polydispersity and the deformability of the particles, or effects related to sedimentation and hydrodynamics, albeit the latter has been very recently ruled out~\cite{fiorucci_effect_2020}.

\subsection{\label{sec:tools}The tools at our disposal}

What options are available to the computational scientist in order to calculate crystal nucleation rate by means of atomistic simulations? To start with, so-called ``brute force'' simulations constitute the most straightforward option: the idea is to bring the simulated liquid (or solution) into a supercooled (or supersaturated) state and simply run a set of unbiased simulations in the hope that the system will crystallize on a reasonably short timescale. At that point, methodologies such as the mean first passage time (MFPT) can be used to extract the nucleation rate~\cite{lundrigan_test_2009,nicholson_analysis_2016}. Brute force simulations are regarded as useful benchmarks to compare the results of enhanced sampling simulations with. However, not very many system will in fact crystallize quickly enough, under experimentally relevant conditions, for this method to be applied. The reality is that, in most cases, enforcing spectacularly strong supercooling or supersaturation conditions is a necessity - and at that point, one should start questioning whether the nucleation events we are seeing do in fact conform to the definition of nucleation as a rare event, an aspect we will address in more detail in Sec.~\ref{sec:rare}. More often than not, brute force simulations aimed at calculating crystal nucleation rates are used for model systems such as LJ liquids or colloids~\cite{baidakov_spontaneous_2019,ouyang_entire_2020}. However, particularly impressive results have been obtained of late for metallic alloys as well~\cite{an_linear_2018,mahata_understanding_2018}, in some cases via simulations involving millions of atoms~\cite{shibuta_heterogeneity_2017}.

In order to compute crystal nucleation rates for realistic systems or indeed for model systems within a wider range of supercooling or supersaturation, enhanced sampling techniques represent the tools of the trade. These can be classified as free energy-based or path sampling-based techniques. The former seeks to obtain information about the thermodynamics of the process, and as such they do not offer immediate access to the kinetics of nucleation. However, it is possible, in principle, to use the outcomes of e.g. umbrella sampling ~\cite{leyssale_molecular_2003,punnathanam_crystal_2006,yi_molecular_2009,filion_crystal_2010} or metadynamics simulations~\cite{quigley_metadynamics_2008,quigley_metadynamics-based_2009,giberti_metadynamics_2015,niu_temperature_2019} (most prominently, the free energy profile or surface) as the starting point to compute the nucleation rate. Popular options in this context include the family of  Bennett-Chandler methods for the calculation of rate constants~\cite{menzl_s-shooting_2016} and the approach recently pioneered by Tiwary and Parrinello~\cite{salvalaglio_assessing_2014,salvalaglio_overcoming_2016}.

In contrast, path sampling-based techniques such as transition path sampling (TPS)~\cite{moroni_interplay_2005,arjun_molecular_2021,bolhuis_transition_nodate}, transition interface sampling (TIS)~\cite{van_erp_elaborating_2005,valeriani_rate_2005,arjun_rate_2020} and forward flux sampling (FFS)~\cite{berryman_sampling_2010,becker_non-stationary_2012,haji-akbari_direct_2015,sosso_microscopic_2016,sosso_unravelling_2018} all allow for the direct calculation of nucleation rates. Interestingly, until quite recently not many examples of any of these methodologies being applied to the calculation of $\mathcal{J}$ could be found in the literature, most likely due to the high computational cost associated with converging these algorithms. However, the ever-growing capabilities of high performance computing facilities, together with the trivially parallel nature of the simulations often needed to sample the crystallization paths, have massively boosted the feasibility of path-sampling methods in the last few years. 

Perhaps unsurprisingly, ML has started to help in the context of enhanced sampling methods as well - not so much as to drive the nucleation process itself but to aid and complement the structural analysis of the newborn nuclei as well as the nucleation paths. Recent examples include the work of Desgranges and Delhommelle~\cite{desgranges_crystal_2018}, where ML is used to infer the free energy of a LJ system over a wide range of densities and temperatures; the work of Bonati and Parrinello~\cite{bonati_silicon_2018}, in which ML is used to train a model capable of describing the crystal nucleation of silicon, starting from a dataset obtained by metadynamics simulations; and the recent work of Adorf \textit{et al.}~\cite{adorf_analysis_2020} where ML has been harnessed to analyze the nucleation pathways of colloidal systems. 

Lastly, seeding techniques~\cite{knott_homogeneous_2012,espinosa_calculation_2016,separdar_molecular_2021,sun_overcoming_2018} can also be used as a starting point to extract crystal nucleation rates. These approaches relies on sampling the evolution of sets of crystalline nuclei of a given size that are inserted into the supercooled liquid or supersaturated solution. Particularly famous examples include the nucleation of ice~\cite{goswami_seeding_2020} and NaCl~\cite{zimmermann_nucleation_2015}.

\subsection{\label{sec:force}The usual suspects: classical nucleation theory and force fields}

The inconsistencies between experimental and simulated crystal nucleation rates, such as those we have discussed in the case of colloidal systems in Sec.~\ref{sec:comp}, are often thought to be due to the usage of CNT and/or a particular force field. The two are not unrelated either, in that several thermodynamic parameters needed to estimate $\mathcal{J}$ via CNT, such as the chemical potential difference between the liquid and the crystal, $\Delta\mu$, and crystal-liquid interfacial free energy $\gamma$, vary wildly according to the choice of a given force field. Several attempts to improve upon the value of $\Delta\mu$ provided the force field in question can be found in the recent  literature. For instance, Wang \textit{et al.}~\cite{wang_classical_2020} have used a second-order Gibbs-Thompson equation to obtain a reliable description of the temperature dependence of $\Delta\mu$ in the case of ice nucleation. In other studies, the concept of an ``effective'' $\gamma$ has been adopted in, e.g.  Ref.~\citenum{zimmermann_nucleation_2015} to describe the nucleation of NaCl from solution, instead of the often-used value of $\gamma$ referring to a macroscopic flat interface. In fact, evidence from the field of ice nucleation confirms that values of $\gamma$ for finite nuclei extracted from simulations via seeding \cite{espinosa_seeding_2016} or umbrella sampling \cite{doi:10.1063/1.3677192} are indeed quantitatively different from those computed from planar interfaces~\cite{limmer_phase_2012,espinosa_icewater_2016,ambler_solidliquid_2017}. Taking into account the Tolman correction~\cite{joswiak_size-dependent_2013} contributes (partially) to bridging the difference, noting that the Tolman \emph{length} encapsulates  only the first term in an expansion of a difference which may not be small.

Indeed, a major point of contention with respect to CNT is its usage at the microscopic scale, where the distinction between solid and liquid phase is not clear cut and the actual value of the thermodynamic parameters involved is bound to differ from their macroscopic counterparts. Even when applying this correction, however, discrepancies of about fifteen orders of magnitude with respect to $\mathcal{J}$ can still be found in Ref.~\citenum{zimmermann_nucleation_2015}. The Joung-Cheatham force field~\cite{joung_determination_2008} for NaCl and SPC/E water used might be partially responsible for that discrepancy still, the authors suggest - but it has to be said that choice is superior to that adopted in previous works where the GROMOS85 force field, which overestimates the stability of the NaCl wurtzite structure, was used~\cite{giberti_transient_2013}. In Ref.~\citenum{zimmermann_nucleation_2015} it is acknowledged that no current force field for NaCl will perfectly capture the chemical potential difference between ions in solution and ions encapsulated in a bulk crystal and hence reproduce exactly the experimental solubility. However, it is argued that provided one compares between simulation and experiment at the same value of this chemical  potential difference (but different absolute concentration) one should be able to draw meaningful conclusions. This requires the crystal solubility within the chosen force field to be known accurately. In the case of the Joung-Cheatham model various estimates of this solubility have been made~\cite{moucka_molecular_2011,aragones_solubility_2012,mester_temperature-dependent_2015}
eventually reaching a consensus~\cite{benavides_consensus_2016} subsequently confirmed by increasingly advanced methods~\cite{boothroyd_solubility_2018} to be approximately half of the experimental value. This indicates that force fields tractable to nucleation simulations are far from sufficiently accurate in absolute terms. Later work \cite{zimmermann_nacl_2018} reexamined the results of Ref.~\citenum{zimmermann_nucleation_2015} in light of the revised estimates of model solubility, resulting in much improved agreement with experimental nucleation rates \textit{when compared at equivalent chemical potential difference}. That is, the authors found that reliable results in terms of $\mathcal{J}$ can be obtained when the driving force for nucleation, $\Delta\mu$, and the solubility are consistent with each other and accurate with respect to the particular model/force field used. This means that, even if $\mathcal{J}$ is in agreement with the experiments,  the actual solubility of the force field is, in this case, not consistent with the experimental value at that particular supersaturation. This work also demonstrated a very strong sensitivity to the choice of order parameter used to quantify the nucleus size. We return to this question in section \ref{sec:order}. 

To illustrate how sensitive the calculation of $\mathcal{J}$ is to the inaccuracies of the force field, we refer to the work of Haji-Akbari and Debenedetti~\cite{haji-akbari_direct_2015}, where the ice nucleation rate has been computed via FFS using the TIP4P/Ice water model at the very strong supercooling of 42 K. This choice might appear rather extreme but it is not unusual: when dealing with realistic and/or computationally expensive force fields, resorting to very strong supercooling or supersaturation is often the only way to gain any insight into the nucleation process, even when using state-of-the-art enhanced sampling methods. This is important because in this scenario, one has to extrapolate the value of $\mathcal{J}$ at milder supercooling in order to be able to compare the computational result to the experimental reality. Once again, then, CNT comes into play, and with it multiple potential sources of inaccuracies. In the case of Ref.~\citenum{haji-akbari_direct_2015}, the authors observe a discrepancy of about eight orders of magnitude between simulated and experimental nucleation rates - a difference that can be explained, they argue, by noting that the TIP4P/Ice water models yields a value of $\Delta\mu$ which corresponds to an enthalpy difference about 20\% smaller than the experimental value. Similar arguments can be found in other recent works. For instance, Arjun and Bolhuis~\cite{arjun_unbiased_2019} put forward the usage of a specific water model as the reason why the free energy barriers relative to the nucleation of methane hydrates, obtained via TPS, are found to be lower with respect to previous results.

Indeed, the reliability of water models is a major issue when dealing with crystal nucleation from solution. We have already mentioned the case of NaCl, but perhaps the most prominent scenario in this context is that of biomineralisation, where the water model has to capture the complex interaction with the mineral under investigation. In their recent review, Demichelis \textit{et al.}~\cite{demichelis_simulation_2018} argue that an accurate description of the solubility of the mineral, a quantity often neglected in the first parametrisations of force fields for simulations of biomineralisation, is paramount to obtain robust results. The additional layer of complexity in that field is the scarcity of experimental data, particularly in terms of clustering and speciation (under conditions accessible by both experiments and simulations) available for the computational scientists to build their force field upon. Electronic structure calculations have been used to fill that gap, which in turn highlighted the absolute need for polarizable force fields when dealing with the crystal nucleation of biominerals~\cite{demichelis_simulation_2018}. Unfortunately, the additional computational expense of such models makes quantitative calculation of $J$ intractable at this time.

Another example, yet again from the ice nucleation field, is given by the recent work of Shi and Tanaka~\cite{shi_homogeneous_2019}, where the authors have found that in the case of the TIP5P model the treatment of electrostatic interactions has a huge impact on the nucleation process, to the point where results massively favoured a specific (ferroeletric) crystalline phase. The fact that truncating electrostatic or indeed even non-bonded interactions can have a substantial effect on the property of the force field - and thus, indirectly, on the estimate of $\mathcal{J}$ - is well known (see e.g. Ref~\citenum{fitzner_communication_2017}) but not necessarily discussed in a transparent, reproducible manner in all computational studies so far. In the same work~\cite{shi_homogeneous_2019}, Shi and Tanaka also point out that while CNT assumes the liquid to be a perfectly homogeneous phase, this might not be the case.

The fact that supercooled liquids and supersaturated solutions exhibit both structural and dynamical heterogeneities is well established, but only in the last few years have we started to witness the first attempts to make a connection between the pre-ordering of the liquid and crystal nucleation. As an example, Menon \textit{et al.}~\cite{menon_role_2020} have recently investigated the crystal nucleation of molybdenum via TIS simulations, finding that the emergence of crystal-like precursors plays a role in polymorph selection. As it concerns dynamical properties, the work of Fitzner \textit{et al.}~\cite{fitzner_ice_2019} has recently established a link between the dynamical heterogeneity of supercooled liquid water and the occurrence of ice nucleation.

In summary, we hope we have convinced the reader that the accuracy of the force fields we use is intertwined with the reliability of our CNT predictions. However, we argue that this picture is far from complete, and that there exist several additional issues that might come into play when computing $\mathcal{J}$ via molecular simulations. In the next section, we will highlight seven such aspects, presenting some relevant examples from the recent literature with the aim to bring together the efforts of the community toward an ever-increasingly accurate picture of the crystal nucleation process from a microscopic perspective.

\section{\label{sec:seven}The Seven Deadly Sins}

\subsection{\label{sec:rare}Rare events - or are they?}

The central idea at the heart of the CNT kinetics was developed by Becker and D\"{o}ring in 1935~\cite{becker_kinetische_1935,jackson_nucleation_2005}: the time evolution of the distribution of crystalline clusters is treated via a formalism equivalent to that of chemical rate equations - under a number of assumptions. For instance, they assumed that the nuclei either shrink or grow via losing or gaining a single atom, particle or molecule to the existing nucleus. Most relevant to this section, though, is the assumption of a quasi-stationary distribution (QSD) of nuclei~\cite{lutsko_classical_2013}, which is supposed to not get depleted in time by the formation of critical nuclei - once critical clusters form, they are removed from the distribution and new, smaller nuclei are added, so as to achieve a steady state. Provided that the free energy barrier associated with the formation of critical nucleus is ``high enough'', the supercooled liquid (or, the supersaturated solution) is to be found in a metastable state with respect to the crystalline phase within a time scale much longer than its relaxation time, thus allowing us to compute $\mathcal{J}$ as a steady-state crystal nucleation rate.

In this scenario, we are dealing with a rare event characterised by a survival probability $P_{liq}(t\mbox{*})$for the supercooled liquid or supersaturated solution that decreases exponentially with time $t$ as e.g. $P_{liq}(t\mbox{*}) = \exp(-\mathcal{J}t\mbox{*})$. In other words, the nucleation times (to be observed across a sufficiently large ensemble of either experiment or simulations, given the stochastic nature of the nucleation process) are distributed according to  Poisson statistics. In the context of unbiased simulations of crystal nucleation, verifying this condition is relatively straightforward, as illustrated in Fig.~\ref{FIG_5} where we report the time evolution of 220-820 (see appendix A for further details) MD trajectories of a LJ system at different supercooling and the corresponding survival probability for the liquid phase. At very strong supercooling (Fig.~\ref{FIG_5}a), the free energy barrier is so low that the liquid is basically unstable, as opposed to metastable, with respect to the liquid phase. As such, there is no incubation (or waiting) time, as the system immediately proceeds to crystallise on a timescale comparable to that of its relaxation time. The resulting survival probability is thus a step function which tell us that, in these conditions, we are not dealing with a rare event in the first place. At milder supercooling (Fig.~\ref{FIG_5}b), we reach a ``butter zone'' (i.e. a temperature regime ideally suited to extract, in this case, $\mathcal{J}$) where we can sample $P(t)$ across a sufficiently large time scale, long tails included. This is an ideal setting to accurately compute $\mathcal{J}$. Finally, at very mild supercooling (Fig.~\ref{FIG_5}c) we are obviously dealing with even rarer events - however, the timescale associated with the nucleation process is such that we can only sample a very small portion of the relevant $P(t)$, thus robbing us from the possibility of investigating the actual decay of this function and assessing whether it is truly consistent with a Poisson distribution. 

It is important to point out this ``butter zone'' might simply not exist for some systems, particularly slow-diffusing ones for which the balance between supercooling (or supersaturation) and atomic/molecular mobility results in time scales too long to be probed by means of unbiased simulations even at strong supercooling. The opposite problem, that of ``not rare enough'' events, is also encountered experimentally and it might coincide with the onset of spinodal decomposition. An intriguing take on this matter is given by Sear~\cite{sear_what_2016}, who provocatively puts forward the idea that ``the nucleation rate may not exist'', given that, by looking at the experimental data, the survival probabilities measured for a variety of systems are far from being exponential. 

This is especially true when considering heterogeneous nucleation, where structural or even chemical changes in the impurity involved with the nucleation process often result in time-dependence nucleation regimes as opposed to the steady state $\mathcal{J}$ we are after. Fortunately, several studies aimed at rationalising the experimental data available to us are can be found in the recent literature. As an example, Maggioni and Mazzotti~\cite{maggioni_stochasticity_2017}
have looked into an extensive set of literature data on the crystallization of p-aminobenzoic acid in three different solvents, developing a statistical analysis that allows to quantify the uncertainty of experimental nucleation data - a crucial aspect to facilitate the comparison with simulated nucleation rates. 

An interesting aspect that unifies the majority of the modelling of nucleation data is the assumption that nucleation can be considered as a Markovian process - that is, there is no history dependence. However, it is becoming apparent that this assumption might not always hold. For instance, Jungblut and Dellago~\cite{jungblut_caveats_2015} have found that the nucleation dynamics of a particular LJ system shows non-Markovian aspects due to the lack of a good enough reaction coordinate, an aspect we will discuss in greater detail in Sec.~\ref{sec:order} and that results in underestimating the nucleation rate if using MFPT methods. Indeed, the recent work of Kuhnhold \textit{et al.}~\cite{kuhnhold_derivation_2019}, also investigating the crystal nucleation of a LJ melt, highlighted the emergence of non-Markovian dynamics and went as far as saying that CNT has to be considered as the limiting case of a more general theoretical framework that includes history-dependent aspects. In the words of the authors, crystal nucleation may be ``neither Markovian nor diffusive''. 
It is important to point out that the term ``non-Markovian'' can refer to two distinct issues, both of which will be discussed in this work. Firstly, the occurrence of nucleation events itself might or might not be a Markovian process. But then, there is the question of whether the \textit{description} of the crystallization process, by means of the evolution of the order parameters we shall discuss in Sec.~\ref{sec:order}, is Markovian or not. 

At this stage, we should ask ourselves how would enhanced sampling fare in extracting accurate nucleation rates, given that even unbiased simulations struggle in some cases to recover reliable estimates of $\mathcal{J}$. A central argument in this context is quantifying the uncertainty associated with any given methodology. The discussion contained in Ref.~\citenum{angulo_new_2016} highlights once more the importance of sampling a QSD: in particular, the work of Binder \textit{et al.}~\cite{binder_generalized_2015} on parallel replica dynamics offers some practical considerations that allow the computational scientist to analyse the obtained nucleation times. If the latter are exponentially distributed, we \textit{might} be working in conditions close enough to a QSD - but not necessarily. Cases where the $P(t)$ is exponential but the Markov process did not reach the QSD are still a possibility. Similar arguments apply to the case of extracting nucleation rates starting from the ``infrequent metadynamics'' framework of Tiwary and Parrinello~\cite{tiwary_metadynamics_2013}. In particular, Salvalaglio \textit{et al.} have devised a simple and computationally inexpensive methodology, based on the Kolmogorov–Smirnov test, to assess the reliability of the kinetics of nucleation obtained from metadynamics simulations. This framework identifies several problematic cases where e.g. the applied bias is strong enough for the system to move away from the QSD~\cite{salvalaglio_assessing_2014,khan_how_2020,bal_nucleation_2021}. 

\begin{figure*}[htbp]
\begin{centering}
\centerline{\includegraphics[width=0.8\textwidth]{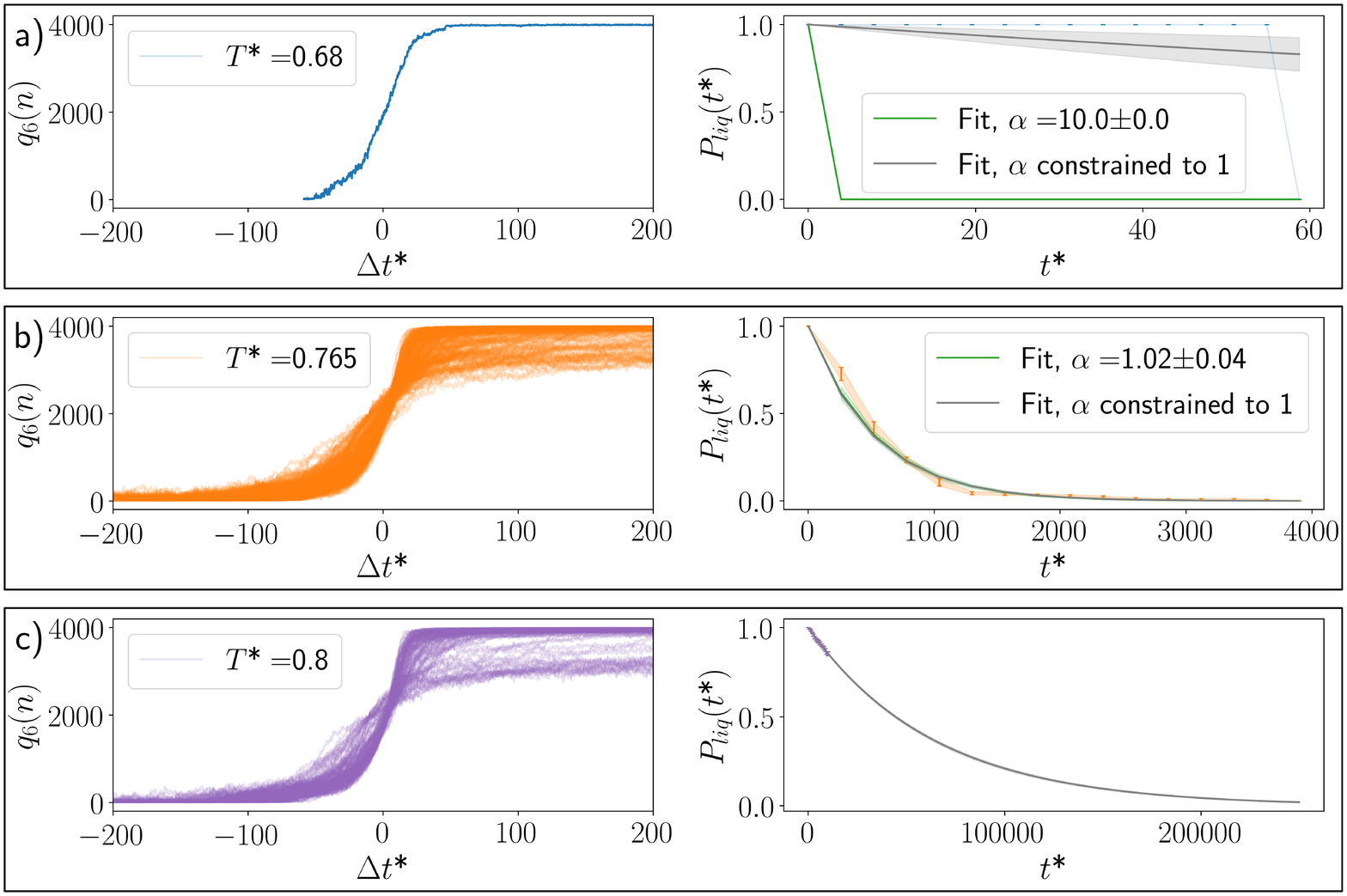}}
\par\end{centering}
\protect
\caption{\textit{Crystal nucleation of a LJ liquid at different supercooling.}  The freezing curves (left panels, shifted so that the nucleation times lie at $\Delta \, t\mbox{*}=0$) and survival probabilities $P_{liq}(t\mbox{*})$ (right panels, as a function of time elapsed, $t\mbox{*}$) for the same LJ system at different (reduced) temperatures $T\mbox{*}$. a) Strong supercooling ($T\mbox{*}=0.68$): the liquid is unstable with respect to the crystalline phase. b) ``Butter zone'' ($T\mbox{*}=0.765$): it is possible to sample the exponential decay of $P_{liq}(t\mbox{*})$ across a sufficiently long time scale. c) Mild supercooling ($T\mbox{*}=0.8$): it is possible to accumulate statistics with respect to the nucleation times only within a very limited time scale compared to the decay of $P_{liq}(t\mbox{*})$. We have fitted the survival probabilities according to the CNT-like expression $P_{liq}(t\mbox{*}) = \exp[-(\mathcal{J}t\mbox{*})^{\alpha}]$, where $\alpha$ is a fitting parameter accounting for the possibility of non-exponential nucleation kinetics. The uncertainty associated with the unconstrained fit is shown as shaded regions. The details of the LJ simulations are given in appendix A.}
\label{FIG_5}
\end{figure*}

\subsection{\label{sec:size}Finite size effects}

The fact that finite size effects can have a substantial impact on the estimate of $\mathcal{J}$ should hopefully come as no surprise. A frequently encountered rule of thumb consist in working with a simulation box larger than two times the extent the critical nucleus size in any given dimension. However, the reality is that, without a thorough investigation of finite size effects using e.g. increasingly large simulation boxes, the extent of this effect is impossible to quantity \textit{a priori}. This is a real issue, as in many cases one can rarely afford the computer time needed to repeat their simulations with larger simulation boxes - particularly when computing nucleation rate for realistic systems. In addition, the impact of finite size effects is related to the supercooling or supersaturation conditions as well. The case of e.g. mild supercooling is very well-known: as the critical nucleus size becomes (exponentially) larger as the supercooling decreases, larger boxes are, by definition, a necessity. However, at strong supercooling, the average density of pre-critical nuclei is much higher than what is observed at mild supercooling, thus posing the problem of the interaction between different, smaller nuclei (as opposed to the very rare / large critical nuclei observed at mild supercooling). Finite size effects are also strongly connected to the issue of solute depletion we will discuss in Sec.~\ref{sec:solute}.

Attempts to quantify the severity of finite size effects in simulations of crystal nucleation date back to the seminal work of Honeycutt and Andersen on LJ systems in 1986~\cite{honeycutt_small_1986}, where the density of critical nuclei was investigated. The actual influence of finite size effects on the calculation of $\mathcal{J}$ has been quantified much later, though - the work of Huitema \textit{et al.} in 2000 being a excellent example~\cite{huitema_thermodynamics_2000}. By then, the typical system size was in the region of 10$^4$ LJ particles. These days, however, in many cases we can afford to investigate much larger systems. For examples, Ouyang \textit{et al.} have recently probed the finite size effects on the crystal growth rate of LJ systems using models containing up to 10$^6$ LJ particles~\cite{ouyang_entire_2020} , taking advantage of GPU-accelerated molecular dynamics simulations.

\begin{figure}[htbp]
\begin{centering}
\centerline{\includegraphics[width=0.6\textwidth]{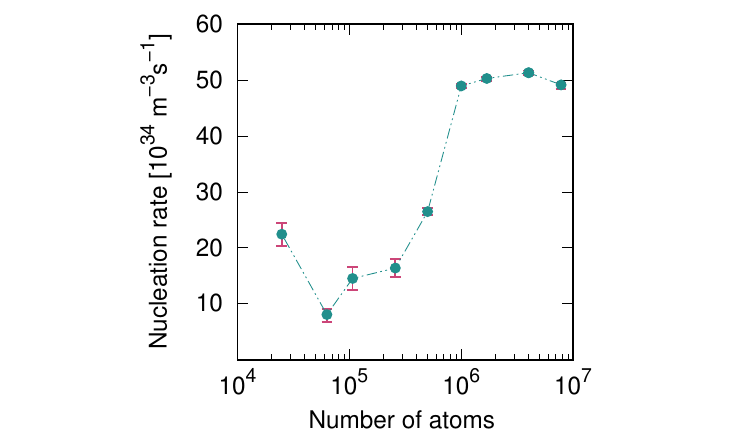}}
\par\end{centering}
\protect
\caption{\textit{The impact of finite size effect on the estimate of crystal nucleation rates.} Nucleation rate for atomistic models of Al, obtained via mean the first-passage time (MFPT) method, for different system sizes. These simulations took advantage of the computational efficiency of the second nearest-neighbor modified embedded atom method (2NN-MEAM) interatomic potential to explore finite size effects arising when dealing with up to several millions of atoms. Adapted from Ref.~\citenum{mahata_size_2019}}
\label{FIG_2}
\end{figure}

Perhaps one of the most striking results within the recent literature concerning finite size effects in the context of crystal nucleation rates is the work of Mahata and Zaeem~\cite{mahata_size_2019}, which focuses on the crystal nucleation of elemental metals described via the second nearest-neighbor modified embedded atommethod (2NN-MEAM) interatomic potential. As illustrated in Fig.~\ref{FIG_2}, the authors have considered models containing up to 10$^7$ atoms, providing reliable estimates of $\mathcal{J}$ as a function of system size. While many would probably argue that discrepancies barely spanning a single order of magnitude are to be considered as small in the context of $\mathcal{J}$ estimates, it is intriguing to observe that finite size effects are still very much present even in systems containing millions of atoms. In addition, the results reported in Fig.~\ref{FIG_2} show a non-monotonic dependence of $\mathcal{J}$ with respect to the system size. This is an intriguing finding, in that it is rather common to assume that the presence of finite size effects tends to consistently overestimate the nucleation rate. However, it appears the picture is more nuanced - and, certainly largely unexplored still. Indeed, the recent work of Hussain and Haji-Akbari~\cite{hussain_how_2021}, which thoroughly explored the impact of finite size effects in the context of simulations of heterogeneous ice nucleation, shows a positive, linear correlation between the ($\log_{10}$ of) the nucleation rate and the system size. The same work provides a number of practical guidelines, particularly in terms of quantifying the spurious interactions between pre-critical nuclei due to small simulation boxes and how to differentiate such interactions from those originating from very strong supercooling regimes. These concepts can and should be considered when dealing with simulations of homogeneous nucleation as well.

An interesting aspect of finite size effects is that the latter are almost always thought of in terms of structural correlations within the supercooled liquid or supersaturated solution. However, recent evidence suggests that dynamical correlations might also play a role (see the excellent review of Zanotto and Montazerian~\cite{zanotto_dominant_2020}). For instance, the work of Fitzner \textit{et al.}~\cite{fitzner_ice_2019} established a correlation between dynamical heterogeneities within supercooled liquid water and the emergence of ice nuclei, and similar findings have been reported for a diverse portfolio of systems, from LJ liquids~\cite{pedersen_how_2021} to oxides~\cite{gupta_role_2016,abyzov_effect_2017} and metallic glass formers as well~\cite{puosi_nucleation_2019}. Dynamical correlations can span much larger length scales than structural ones, thus prompting the question of whether up-to-now undetected finite size effects in terms of dynamical properties might be present in our simulations of crystal nucleation - a largely unexplored possibility. 

Finally, finite size effects are important when computing the solubility of e.g. ions in solution~\cite{espinosa_calculation_2016}, which in turn is key to calculate $\mathcal{J}$ if leveraging CNT via, for instance, seeded approaches. The case of NaCl is especially relevant, and will thus be discussed in the next section, where we will tackle the emergence of solute depletion effects - an occurrence very much related to the size of the simulation box and thus to the possibility of finite size effects as well.

\subsection{\label{sec:solute}Solute depletion}

\begin{figure*}[htbp]
\begin{centering}
\centerline{\includegraphics[width=0.9\textwidth]{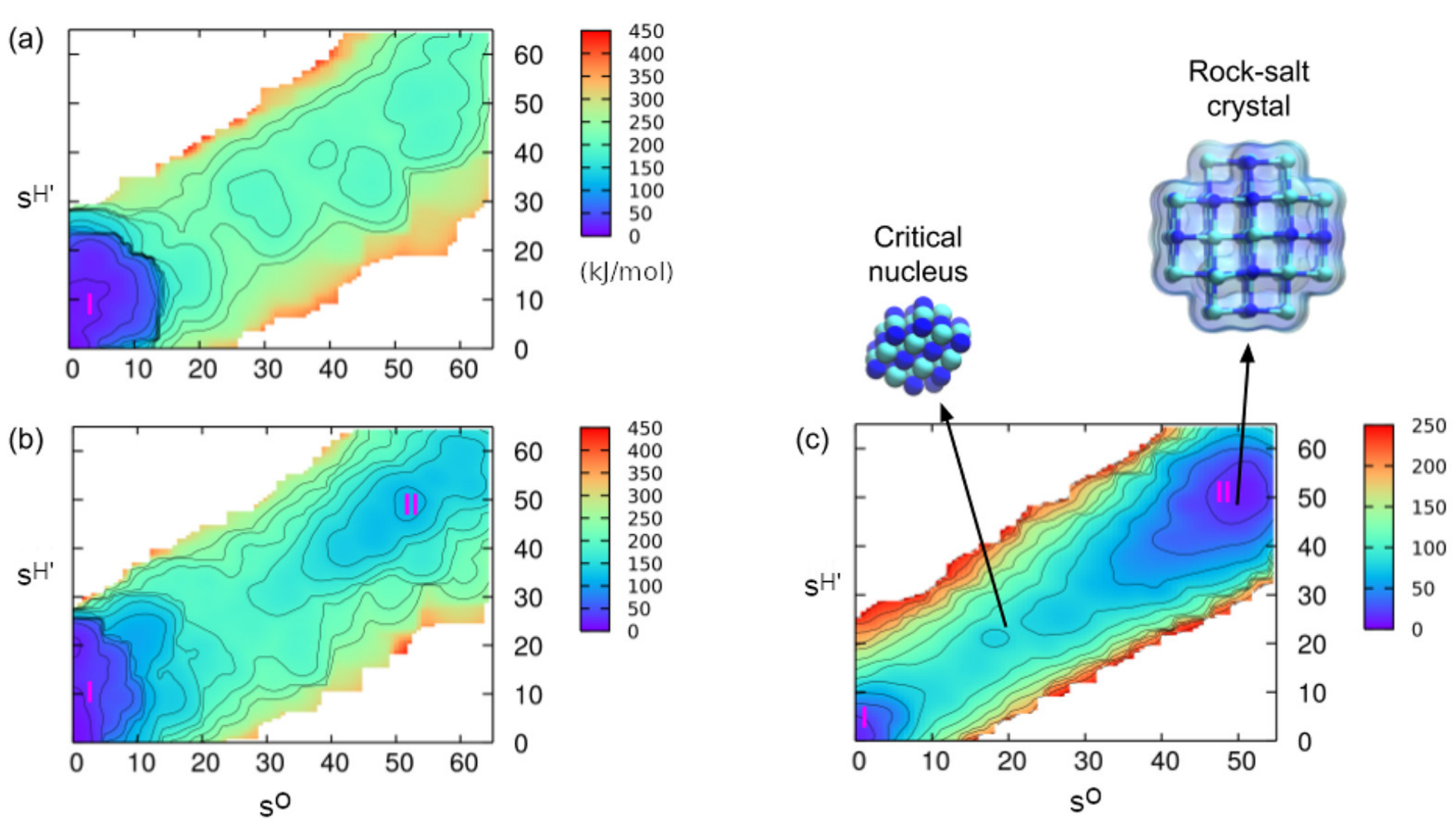}}
\par\end{centering}
\protect
\caption{\textit{Simulating the crystal nucleation at constant chemical potential.}  Free energy surfaces, obtained by means of metadynamics simulations, relative to the nucleation of NaCl cystals from solution. The two order parameters $S^O$ and $S^{H'}$ provide a measure of the crystalline order and the hydration number of Na$^+$ ions, respectively. (a) NVT ensemble; (b) Constant chemical potential via the C$\mu$MD method discussed in Ref.~\citenum{karmakar_molecular_2019}; (c) Same as (b) but at higher supersaturation. Note that the conventional NVT setup fails to identify the basin corresponding to the crystalline phase. Adapted from Ref.~\citenum{karmakar_molecular_2019}}
\label{FIG_3}
\end{figure*}

The molecular simulator is usually interested in quantifying the nucleation rate by by studying a small volume of the system, representative of the bulk metastable parent phase. Simulations should ideally capture the coupling of this volume to its environment (i.e. the rest of the bulk system) via exchange of both particles and heat. Most simulations of nucleation from solution have been conducted with a constant number of solute particles. The nucleation of a solid phase when working with relatively small simulation boxes leads to a change in the effective supersaturation of the simulation due to the fact that the crystalline phase nucleates at the expenses of the solution phase, thus depleting the latter to an extent proportional to the number of particles/atoms/molecules contained in the simulation box. Real systems are large enough to compensate the resulting change in terms of chemical potential, but in atomistic simulations this is a serious issue, which is less evident (albeit still present and, we argue, rather unexplored up to now) for crystal nucleation from a supercooled liquid (where as opposed to ``depletion'' the emphasis is on the density difference between the liquid and the crystal) but quite spectacularly evident for crystal nucleation from a supersaturated solution. As an example, in Fig.~\ref{FIG_3}a it can be seen the free energy surface relative to the formation of crystalline NaCl from solution simply lacks the basin corresponding to the crystalline phase when performing simulations within the NVT ensemble. 
It is important to note, however, that solute depletion is not a prerogative of simulations and it can occur in actual nucleation processes as well. Depleted regions around the growing crystalline nuclei are also present in reality: they might be replenished quickly enough to conform to the CNT assumption of a constant background supersaturation but they might also be replenished very slowly (depending on e.g. the diffusivity of the solute) - a situation closer to simulations but not consistent with CNT.

A number of different approaches to circumvent this long-standing issue have been proposed~\cite{sosso_crystal_2016}. As straightforward as it might sound, simply increasing the size of the simulation box might be a viable option. For instance, Zimmermann \textit{et al.}~\cite{zimmermann_nucleation_2015} provided practical guidelines as to how big the simulation box should be in order to avoid solute depletion effects when using seeded MD simulations to extract crystal nucleation rates. In particular, the box should be large enough to (i.) contain enough solute to form a critical nucleus, and (ii.) allow, once a critical nucleus has formed, the solution to remain in a supersaturated state. In practice, simulations with fixed numbers of solute particles must be much larger than these lower bounds to be representative of nucleation in a system which can exchange solute with its surrounding bulk environment to replenish concentration in the vicinity of the nucleus. An alternative is to use analytical corrections to the free energy surfaces (to be used as staring points in order to estimate $\mathcal{J}$) obtained via ``closed'' simulations (e.g. NVT or NPT ensembles), as exemplified by the seminal work of Agarwal and Peters~\cite{agarwal_nucleation_2014}. In 2016, Salvalaglio \textit{et al.} computed $\mathcal{J}$ (relative to the condensation of liquid droplets from vapour as opposed to crystal nucleation) for small systems using infrequent metadynamics~\cite{tiwary_metadynamics_2013} and applying a bespoke analytical correction~\cite{salvalaglio_overcoming_2016}. Their results have been very recently validated by Bal \textit{et al.}~\cite{bal_nucleation_2021}.

More sophisticated methodologies often seek to modify the computational setup itself. Recent advances in this context include the work of Liu \textit{et al.}~\cite{liu_modelling_2018} which harnesses the so-called string method in collective variables (SMCV~\cite{maragliano_string_2006}) to work in the osmotic ensemble ($N_{\text{liquid}}$,$\mu_{\text{crystal}}$,P,T). This is an intriguing thermodynamic ensemble which, applied to crystal nucleation, seeks to conserve the number of particles in the liquid phase $N_{\text{liquid}}$ as well as the chemical potential of the crystalline phase $\mu_{\text{crystal}}$, in addition to temperature and pressure. However, it has to be said that, to our knowledge, the SMCV method has never been used to compute $\mathcal{J}$ directly so far - albeit it can be used to compare the relative nucleation rate of two different crystal polymorphs~\cite{maragliano_string_2006} (a very useful feature).

Another approach is to build on multiscale methods such as the adaptive resolution simulation (AdResS) scheme pioneered by Wang \textit{et al.}~\cite{wang_adaptive_2012}. In this case, the system is divided into smoothly connected regions where atoms/particles/molecules are move from an atomistic to a coarse-grained description. In doing so, the balance of the different degrees of freedom involved with the atomistic and coarse-grained regions allows one to work in what is effectively a grand canonical ensemble ($\mu$,V,T). The latest development of this approach can be found in Ref.~\citenum{delle_site_molecular_2019}, but its potential application to the actual calculation of $\mathcal{J}$ remains to be explored.

Crystal nucleation rates have been instead recently computed, once more in the case of NaCl, by Karmakar \textit{et al.}~\cite{karmakar_molecular_2019} using a variant of the constant chemical potential molecular dynamics (C$\mu$MD) method developed by Perego \textit{et al.}~\cite{perego_molecular_2015}. In this case, the underlying framework is that of well-tempered metadynamics~\cite{barducci_well-tempered_2008}, which allows to obtained free energy surfaces such as those illustrated in Fig.~\ref{FIG_3}. In stark contrast with the NVT scenario depicted in Fig.~\ref{FIG_3}, the latest modification of the C$\mu$MD method correctly identifies the free energy basing corresponding to the emerging crystalline phase, as reported in Fig.~\ref{FIG_3}b and Fig.~\ref{FIG_3}c - where the supersaturation has been purposefully increased so as to highlight the stability of the crystal. It is also worth noticing that a modification of the original C$\mu$MD approach which should be able to lower the computational cost of the methodology has recently been put forward by Chen and Ren~\cite{chen_molecular_2020}.

\subsection{\label{sec:thermo}Microscopic kinetics}

Solute depletion represents one limitation of attempting to approximate an open system (coupled to its surrounding bulk parent phase) with a closed simulation. As well as exchange of particles with its surroundings, a real system can also exchange heat. This is relevant to simulation of nucleation from the melt as well as from solution. 

In molecular dynamics simulations, exchange of heat with implicit surroundings is accomplished by augmenting equations of motion with stochastic and dissipative forces, or by augmenting the system Hamiltonian with additional degrees of freedom which accurately mimic the effect of coupling to a heat bath for the purposes of sampling.  The resulting microscopic dynamics are however entirely fictitious. It is common to ``thermostat'' all degrees of freedom in this way.

Consider for example a growing nucleus. As the parent phase transforms into a crystal, latent heat is released at the interface. In a standard molecular dynamics simulation this heat is immediately removed via the thermostat on a (typically) short timescale determined by the thermostat coupling parameter, such that the temperature of the system remains both uniform and constant throughout the simulation. Similarly a shrinking crystal will absorb heat from the parent phase which is then rapidly replenished via the thermostat.

Thermostatted molecular dynamics simulations of this kind are entirely appropriate to calculations of the free energy barrier to nucleation  within CNT.
The ensemble of nuclei which have size $n$ is assumed within CNT to contain both shrinking and growing members such that the net flow of heat between nuclei and surroundings is zero, as is the case for an ensemble of configurations sampled via a simulation in which all degrees of freedom are coupled to a thermostat. This does however imply that techniques which infer the free energy gradient from the average dynamics of the nucleus size metric at each $n$~\cite{knott_homogeneous_2012} must sample over ``swarms'' of trajectories which contain both growing and shrinking nuclei.  

However for ``brute force'' or rare event methods it is not immediately clear that thermostatting all degrees of system is the appropriate choice. Consider again a growing nucleus releasing heat into its immediate surroundings. In a real/open system this heat must be transported away from the nucleus before reaching the implicit heat bath represented by the surrounding bulk parent phase. This may occur on a timescale slower than the growth rate of the nucleus, meaning that subsequent additions to the nucleus take place at an elevated local kinetic temperature. This entirely physical effect would not be captured via a simulation in which a thermostat is applied globally. This has been discussed in the context of crystal growth at planar ice/water interfaces~\cite{rozmanov_temperature_2011} as an explanation for lower growth rates when compared to experiment, but not explored quantitatively in simulations of nucleation to our knowledge. This is most likely to be most relevant at strong supercooling where rapid crystal growth is hindered by the limited speed at which heat can be transported away from the interface.

Within CNT, the rate of attachment to a critical nucleus appears in the exponential prefactor. The use of a global thermostat is therefore unlikely to have an impact on nucleation rates calculated from simulation in the case of freezing from the melt where uncertainties typically span several orders of magnitude.

In the case of nucleation from solution, the (additional) analogous problem is transport of solute to/from the nucleus. Indeed simulating with a constant number of solute particles (as is common) is analogous to having no thermostat at all. A growing nucleus will deplete the surrounding region of solute, creating a region of effective lower supersaturation which in turn will necessitate a larger critical nucleus. The crystal can grow only until the surrounding region is no longer supersaturated with respect to the growing phase and an equilibrium is reached, preventing the critical nucleus size from ever being reached. A much more serious problem. In fact, one could argue that these limitations in terms of transport affect simulations of nucleation in a very similar fashion to what happens when constraining the number of particles in the system (see Sec.~\ref{sec:solute}).

As discussed above, implementing grand canonical molecular dynamics simulations is increasingly feasible but still unusual. With the sole exception of Ref.~\citenum{perego_molecular_2015}, there has been (to our knowledge) no study of how the location of the insertion/removal region can impact upon growth rates. Ideally this should be optimised in terms of distance from the nucleus and insertion/removal frequency such that the diffusive kinetics of solute movement in to and out of the simulated areas are accurately captured. For example, a solute/solvent force field which accurately captures the solute diffusion coefficient, plus insertion sufficiently far from the nucleus such any memory of the insertion position/orientation of the solute is lost before it can be incorporated into the nucleus. As with the precise details of the thermostat implementation, this is unlikely to have a major impact on calculated nucleation rates due to the dominance of the exponential term.

In context of Monte Carlo simulations of nucleation in simple lattice models, there has been some investigation into how microscopic kinetics impacts the dynamics of the order parameter or collective variable used to characterise progress along the nucleation process, typically the cluster size. \citeauthor{kuipers_limitations_2010}~\cite{kuipers_limitations_2010} have explored the usual assumption that a collective variable or order parameter (e.g. nucleus size)  evolves according to Markovian dynamics. This appears to be a good approximation in the case where insertion of removal of lattice gas particles can occur at any  position (including directly at the cluster boundary). However a choice more appropriate to modelling solute precipitation on a lattice is to use Kawasaki-type local exchange moves~\cite{kawasaki_diffusion_1966} which capture mass transport to/from the nucleus rather than instantaneous insertion/removal. Here a simple Markovian model was found to be less than optimal.

Similarly, in Ref.~\citenum{lifanov_nucleation_2016} it is shown that the quality of fit to a Markovian model of nucleus size evolution is significantly worse when using diffusive microscopic kinetics instead of allowing particles within or at the boundary of a nucleus to be exchanged directly with a reservoir without including transport. This is clearly very relevant when dealing with seeded simulations. When inferring free energy gradients from nucleus size dynamics by fitting a Markovian model, this may lead to substantial sources of error. In the extreme case, nucleation which is diffusion limited implies no separation of timescales between the dynamics of solute particles and the dynamics of the nucleus size coordinate. In such cases calculations of nucleation rates based on dynamics of that collective coordinate on a free energy landscape are inappropriate and would lead to potentially meaningless results. 

We note here that even though a physical choice of microscopic kinetics may invalidate the assumption that the dynamics of reaction coordinate are Markovian, this does not necessitate that a Markovian model (such as that implied by CNT) is incapable of correctly capturing the nucleation rate provided the thermodynamic assumptions of CNT are satisfied. In other contexts~\cite{krivov_reaction_2013} it has been established that it must be possible to correctly predict rates from a Markovian model of the reaction coordinate dynamics (even if those dynamics are not Markovian) provided the reaction coordinate is chosen to be the committor. See Sec.~\ref{sec:order} for further discussion on optimisation of the reaction coordinate, the committor and its relationship to the nucleus size.

\subsection{\label{sec:cnt}Classical Nucleation Theory}

The known limitations of CNT are many and they have been reviewed extensively elsewhere (see e.g. Refs.~\citenum{karthika_review_2016,smeets_classical_2017,vekilov_nonclassical_2020} for some recent perspectives). However, it is important to stress that new issues related to CNT continue to appear as the community pushes the boundary of atomistic simulations to compute $\mathcal{J}$ for either more complex systems or employing novel enhanced sampling techniques. Two- or even multi-step nucleation, a scenario which is not taken into account by the original formulation of CNT, is an excellent example. In fact, in the last few years both experiments and simulations have reached a point where investigating complex crystallization processes such as biomineralisation and the formation of hydrates is now a possibility. These systems are famous for their hotly debated mechanism of crystal nucleation and growth, which involves multiple steps not necessarily well described by even extended versions of CNT. Recent computational efforts in the field of methane hydrates that seems to validate at least in part the usage of CNT in that context include the work of Arjun and Bolhuis~\cite{arjun_rate_2020}, who have used TIS to compute the homogeneous nucleation rate of methane hydrate at relatively low supersaturations. They find that the rate computed via TIS and the rate obtained via CNT (starting from the free energy barrier they have calculated) are in fact in good agreement, and they provide a thorough analysis of the uncertainties associated with their estimates as well. Various extensions to the original CNT framework have been put forward through the years in order to deal with multi-step nucleation: the one pioneered by Peters~\cite{peters_supersaturation_2011} in 2011 has been recently used as the starting point to describe the homogeneous nucleation of methane hydrates~\cite{lauricella_mechanisms_2017} via non-equilibrium MD.

Another issue with the potential of undermining one of the fundamental assumptions of CNT is the emergence of non-Markovian dynamics. Through the lens of CNT, crystal nucleation can be thought as the time evolution of the distribution of crystalline nuclei within the system, which is usually described via the Fokker-Planck equation as a Markovian (i.e. both stochastic and memory-less) process. This assumption has been repeatedly questioned: for instance, Kuipers and Barkema~\cite{kuipers_limitations_2010} have studied nucleation in the celebrated Ising model and concluded that, when dealing with realistic, diffusive dynamics, the Fokker-Plank equation cannot be used and non-Markovian effects should somehow be incorporated instead. The deviation of the nuclei from the spherical shape assumed by CNT (another point of contention) is also blamed as a potential source of uncertainty in the same work.
A more recent example is the work of Kuhnhold \textit{et al.}~\cite{kuhnhold_derivation_2019}, where the authors argue that CNT can in fact be considered as a limiting case of a more general theory that contains memory and out-of-equilibrium effects: broadly speaking, nucleation can not be considered, according to the authors, as either a Markovian or a diffusive process. Related to the issue of non-Markovian dynamics is the fact that the CNT expression for $\mathcal{J}$ refers to a steady state nucleation rate, whilst in many cases we observe transient nucleation. This problem has been previously reviewed by e.g. Sear~\cite{sear_quantitative_2014}. Here, we mention a recent work by Myint \textit{et al.}~\cite{myint_nanosecond_2018}, focusing on the crystallization of ice VII at high pressure, where some aspects of hydrodynamic have been incorporated in a theoretical treatment that allows to relax the assumption of steady-state nucleation.

Very practical considerations about the reliability of CNT often arise when applying the latter to estimate $\mathcal{J}$ starting from the basic ``ingredients'', that is the kinetic prefactor and the free energy barrier $\Delta G(n*)$ associated with the critical nucleus size $n^*$. In turn, $\Delta G(n*)$ can be obtained from the free energy difference $\Delta\mu$ between crystal and parent phase and the interfacial free energy $\gamma$ between the two. The seeded MD approach we have discussed in Sec.~\ref{sec:tools} provides the perfect platform for this discussion, as it usually relies on several estimates of these ingredients, in some cases computed independently and all of them affected by some degrees of uncertainty. An interesting as well as fundamental question concern the definition of $n$, that is the number of molecules within the crystalline nuclei. This deceptively simple question is not easily answered, as it depends on the choice of a specific order parameter - as we shall discuss in detail in Sec.~\ref{sec:order}. In a recent work, Cheng \textit{et al.}~\cite{cheng_gibbs_2017} argue that the usage of the solid-liquid Gibbs dividing surface is the one choice consistent with the CNT formalism and put forward a methodology, based on combining simulations of planar interfaces and three-dimensional nuclei, to alleviate the uncertainty associated with the estimate of $\Delta G(n*)$ and, thus, of $\mathcal{J}$. Another investigation that shed new light onto the error propagation involved with the calculation of $\mathcal{J}$ in the context of seeded MD is the work of Lifanov \textit{et al.}~\cite{lifanov_nucleation_2016} on nucleation in a multi-species lattice model. In addition to some consideration of non-Markovian dynamics, the authors show that, even when taking great care in computing $\Delta\mu$, seeded MD cannot provide accurate estimates of $\Delta G(n*)$ in the low supersaturation regime. To provide some context, the fact that $\Delta G(n*)$ enters as an exponential in the expression of $\mathcal{J}$ implies that a 43\% error with respect to $\Delta G(n*)$ can lead to an uncertainty of some ten orders of magnitudes in terms of the nucleation rate. 

The last aspect we believe it is worth discussing in the context of CNT limitations is the assumption that the nuclei are supposed to perfectly spherical for every value of $n$. Note that, while this is a very common assumption, it is perfectly possible to rewrite the relevant CNT expressions assuming nuclei shapes other than spherical - albeit the situation becomes complicated if one wants to allow for the occurrence of any arbitrary shape.
Experimental measurements, most prominently on colloidal systems, have demonstrated that crystalline nuclei (particularly around the critical size and/or within the early stages of the crystallization process) are rarely spherical, see e.g. Refs.~\citenum{gasser_real-space_2001,wang_direct_2015}. Simulations have been key to show that deviations from the spherical shape are commonly encountered across a much wider range of systems, particularly during the early stages of the nucleation process as well as strong supercooling, where the critical nuclei are smaller and thus affected to a greater extent by potential deviation from the spherical assumption~\cite{bai_test_2005,diaz_leines_atomistic_2017}. This is relevant for seeded simulations of crystal nucleation as well, as the crystalline seeds are usually built as spherical particles~\cite{espinosa_seeding_2016}. Several modifications to CNT have been proposed throughout the years to account for the deviation of the nuclei from a spherical shape - see e.g. the work of Prestipino \textit{et al.}~\cite{prestipino_systematic_2012,prestipino_shape_2014}. A more recent example is provided by the work of Lutsko~\cite{lutsko_systematically_2018}, which entirely removes  the need for the assumption about the spherical shape in the first place.
We also note here that in the context of nucleation in the Ising model, well-established corrections for fluctuations in nucleus shape as well as effective ``surface energies'' (a concept we have already encountered in Sec.~\ref{sec:force})for small nuclei have been explicitly tested against FFS and umbrella sampling calculations and found to be highly accurate~Cai~\cite{ryu_numerical_2010}. 
    
\subsection{\label{sec:order}The choice of the order parameter}

Every route toward the estimate of $\mathcal{J}$ involves the choice of an order parameter, that is a mathematical object that allows us to identify which atoms, molecules or particles belong the crystalline nuclei. In this work, we are going to adope the nomenclature of Peters~\cite{peters_reaction_2016}, according to which a \textit{collective variable} is any function of the full phase space coordinates, an \textit{order parameter} is a collective variable that distinguishes the typical reactant and product configurations, and the \textit{reaction coordinate} (RC hereafter) is a special order parameter 
that accurately quantifies dynamical progress from reactant to product. It is important to stress that when dealing with complex processes such as crystal nucleation, the ``true'' RC might very well be a high-dimensional set of variables that we have no hope to identify correctly in their entirety. What we can do instead is to coarse-grain the manifold of said variables into a handful of tractable order parameters that we treat as our RC. 

Many order parameters to identify crystalline nuclei build on the famous work of Steinhardt \textit{et al.}~\cite{steinhardt_bond-orientational_1983} and are briefly reviewed in Ref.~\citenum{tribello_analyzing_2017}. Topological order parameters such as the permutation invariant vector (PIV, reviewed in Ref.~\citenum{pietrucci_novel_2020} represent a valid alternative, and of course machine learning has found its way into this particular facet of the field as well, not only to craft new order parameters (see e.g. Refs~\citenum{fulford_deepice_2019,kim_gcicenet_2020}) but also to mine the extensive portfolio of the existing ones in the attempt to pinpoint the combination yielding the best accuracy~\cite{doi_mining_2020}. In a similar fashion, there have been several attempts to automate the choice of the RC in the last few years, albeit none - to our knowledge - applied to crystal nucleation just yet. As an example, Krivov~\cite{krivov_protein_2018} has recently suggested an adaptive optimization of a multivalued RC to be used in the context of protein folding, a long-standing problem involving an incredibly high-dimensional space. Another approach is that of the spectral gap optimization of order parameters (SGOOP) put forward by Smith \textit{et al.} ~\cite{smith_multi-dimensional_2018} and applied to the study of the kinetics of chemical reactions. It is important to note that these examples are still very much relevant to our discussion because it is now clear that, in many cases, we need to step away from the assumption that a single RC, such as the size $n$ of the largest crystalline nucleus within the system, would be sufficient to provide a reliable estimate of $\mathcal{J}$.

In fact, crystal nucleation has been traditionally thought of as a rather one-dimensional problem, partially because $n$ is the one and only RC involved in the CNT framework. However, we now have evidence that this is not the case. Concerning the nucleation of crystal from supersaturated solutions, it is understood that, at the very least, one needs to work in a two-dimensional RC space involving the crystallinity as well as the local density (see e.g. Ref.~\citenum{salvalaglio_molecular-dynamics_2015}). When it comes to nucleation from the supercooled liquid phase, Jungblut and Dellago~\cite{jungblut_reaction_2015} have shown in the case of a LJ system that the choice of $n$ as the only RC can lead to substantial memory effects, which invalidate the assumption of nucleation as a Markovian process and result into vastly different nucleation rates (which they evaluate via TIS). Interestingly, these effects are connected to relative abundance of the two polymorphs, bcc and fcc, within the crystalline seed that have been used to initiate the nucleation events. Indeed, polymorphysm is a very common occurrence that is very difficult to assess \textit{a priori}, as even when the most stable crystal form is known, there is no guarantee that the nucleation process would not involve intermediates characterised by different crystalline structures - again, the LJ system provides a striking example in this sense (see e.g. Ref.~\citenum{desgranges_controlling_2007}), or even result in a long-lived metastable bulk phase. 

\begin{figure*}[htbp]
\begin{centering}
\centerline{\includegraphics[width=0.8\textwidth]{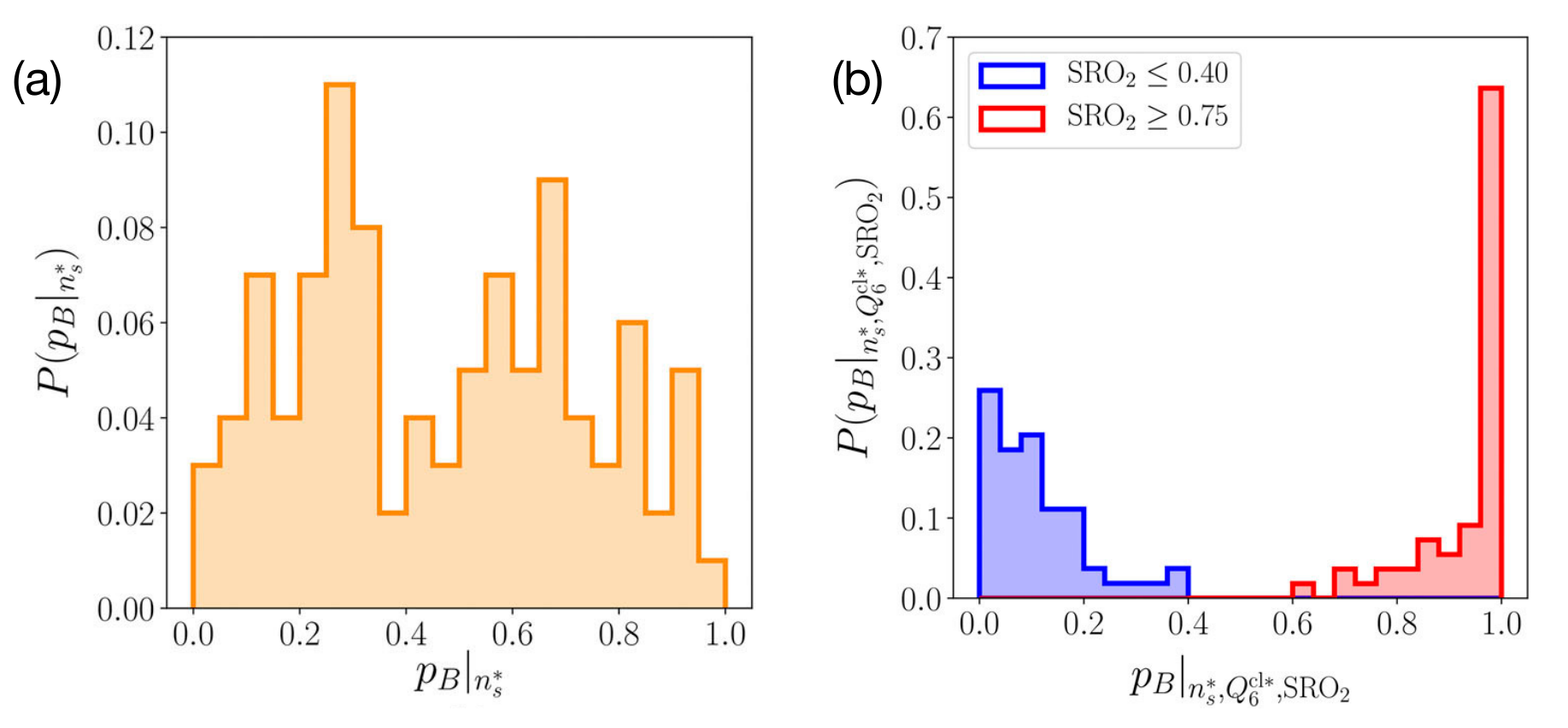}}
\par\end{centering}
\protect
\caption{\textit{Investigating the suitability of order parameters for crystal nucleation.}  (a) Committor distribution $P(p_B|n_s^*)$, where $n_s^*$ corresponds to the number of atoms within the largest crystalline nucleus. (b) Committor distributions $P(p_B|n_s^*,Q_6^{cl*},\text{SRO}_2)$ for two sets of configurations characterised by the same $n_s^*$, the same degree of crystallinity (quantified by the order parameter $Q_6^{cl*}$, inspired by the work of Refs.~\citenum{ten_wolde_numerical_1995,jungblut_pathways_2016}), but different short range order $\text{SRO}_2$. Adapted from Ref.~\citenum{liang_identification_2020}}
\label{FIG_4}
\end{figure*}

An additional layer of complexity has been recently brought to light by Liang \textit{et al.}~\cite{liang_identification_2020}, who have investigated the crystal nucleation molybdenum via TIS simulations. The result of the committor probability $P(p_B|n_s^*)$ they have computed for this system according to the size of the crystalline nuclei alone,  $n_s^*$, is reported in Fig.~\ref{FIG_4}a. Note that the committor distribution fails to yield a well-defined, narrow peak for $P(p_B|n_s^*)$ = 0.5, thus demonstrating the this order parameter does not provide an accurate enough approximation of the reaction coordinate. In fact, not even introducing explicitly an additional degree of freedom in the form of a crystallinity parameter ($Q_6^{cl*}$, inspired by the work of Refs.~\citenum{ten_wolde_numerical_1995,jungblut_pathways_2016})) is sufficient to improve the committor in this case. The authors had to resort to a third order parameter, namely a measure of the short-range order ($\text{SRO}_2$), in order to univocally determine which nuclei are more or less likely to either dissolve into the liquid or proceed toward post-critical sizes. This is illustrated in Fig.~\ref{FIG_4}b, where the committor probability $P(p_B|n_s^*,Q_6^{cl*},\text{SRO}_2)$ is narrowly peaked around zero and one for nuclei characterised by low and high values of $\text{SRO}_2$, respectively, thus indicating that only sufficiently compact nuclei are likely to cross the free energy barrier associated with the nucleation process.

The freezing of water to ice offers an especially challenging testing ground when it comes to the choice of the reaction coordinate. For instance, the work of Prestipino~\cite{prestipino_barrier_2018}, focusing on the thermodynamic of crystal nucleation for a monoatomic water model, shows that the quality of a reaction coordinate cannot be assessed simply on the basis of the free energy barrier height obtained. On the contrary, different choices of different reaction coordinates led to comparable free energy barriers but significantly different fractions of hexagonal ice, cubic ice and ice-0 as well within the post-critical nuclei. Interestingly, in this particular case the committor probabilities for the different reaction coordinates used are quite similar to each other, which implies that none of them alone captures the nucleation process accurately enough. In fact, it has  becoming apparent that in some cases we really need to combine multiple order parameters to obtain a sufficiently accurate reaction coordinate. As an example, Niu \textit{et al.}~\cite{niu_temperature_2019} have recently computed the ice nucleation rate for the TIP4P/Ice model by bringing together a local order parameter with a linear combination of seven long-range descriptors. Hence the relevance of the above mentioned approaches aimed at learn, automate and select multi-dimensional degrees of freedom, as potential avenue to lessen the burden associated with the often employed trial-and-error approach when building the reaction coordinate.

Quantifying the uncertainty associated with the choice of a particular reaction coordinate is also key. In principle, the use of reactive flux calculations (e.g. TIS, FFS) compensate exactly for potential inaccuracy in terms of the reaction coordinate~\cite{peters_reaction_2016}, while
indirect estimates of $\mathcal{J}$ obtained via combining free energy profiles with path-sampling methodologies can be corrected, under certain assumptions, via e.g. a committor analysis. The latter is a popular, albeit computationally extensive, option, and specific flavours are available to identify and mitigate the inadequacy of the reaction coordinate~\cite{peters_reaction_2016,yappert_overdamped_2019}. Systems characterised by a inherently slow dynamics, which we will discuss in greater detail in the next section, are especially challenging to deal with~\cite{beckham_optimizing_2011,lechner_reaction_2011,berezhkovskii_diffusion_2013,krivov_reaction_2013}.

An interesting aspect of relevance for the computational community is the availability of the implementation of the many order parameters we use to construct reaction coordinates - not just as standalone mathematical objects, but in conjunction with MD and MC packages as well, so that they can be leveraged to deploy the enhanced sampling method of choice. In this context, several highly collaborative projects have been born with the aim of facilitate the access to a variety of order parameters for enhanced sampling simulations, such as the SSAGES~\cite{sidky_ssages_2018} suite or the PLUMED consortium the authors are part of~\cite{bonomi_promoting_2019}.

\subsection{\label{sec:slow}Slow dynamics}

Finally, we discuss an assumption implicit in many path-sampling techniques applied to calculation of nucleation rates. Forward-Flux Sampling (FFS) in particular has become the principle workhorse of many rare event studies~\cite{haji-akbari_direct_2015,borrero_simulating_2009}. Alternative techniques such as transition interface sampling (TIS) are also used in the context of nucleation~\cite{arjun_molecular_2021}, but generally acknowledged to be somewhat more complex to implement. The aimless shooting approach to path sampling is a third example~\cite{peters_obtaining_2006}, used in a number of nucleation studies~\cite{peters_reaction_2016}. The relationship between these methods has been very recently summarised in an informative review by \citeauthor{bolhuis_transition_nodate}~\cite{bolhuis_transition_nodate}.

Both FFS and TIS make the implicit assumption that the interval between nucleation events is dominant in determining the nucleation rate. The time taken for the event to occur is neglected. A further practical consideration is that the time taken for the nucleation event to occur must be tractably accessible to simulation. For example, the path-sampling stage of an FFS calculation requires contiguous sampling of trajectories which return to the boundary of the quasi-stationary parent state from an initial configuration containing a critical nucleus. TIS requires sampling of contiguous trajectories which leave the parent state, form a critical nucleus and \emph{then} return to the boundary of the parent state. Many such trajectories must be sampled, in practice limiting the application of these techniques to fast (but still rare) nucleation processes which occur on nano-second timescales or less. 

These two limitations have the effect of limiting the window of supercooling/supersaturation where FFS/TIS calculations can be accurately applied. In freezing from the melt, as supercooling increases the interval between nucleation events becomes smaller, while dynamics become slower. Eventually the assumption of fast nucleation breaks down, as does our ability to sample appropriate trajectories. Applying these methods in regimes where the dynamics of nucleation are slow is likely to be fruitless. 

One approach to obtaining nucleation rates in this regime is to rely on CNT.  The method of \citeauthor{knott_homogeneous_2012}~\cite{knott_homogeneous_2012} (discussed earlier) pioneered the approach of fitting a free energy barrier to gradients inferred from mean growth/shrinkage rates over swarms of short trajectories, precisely to circumvent the need to simulate long trajectories. 
Approaches which do not rely on CNT are also available. The partial path variant of TIS \cite{van_erp_elaborating_2005} requires simulating only short trajectory segments sample an ensemble of trajectories which cross only small windows of the reaction coordinate. The complete nucleation kinetics are subsequently reconstructed from a Markov state model which uses these crossing rates. This relies on a loss of memory between non-adjacent states, i.e. that the dynamics of the the reaction coordinate are Markovian on at least the broad scale corresponding to crossings between windows . This assumption may lead to correct rates even if the true dynamics are non-Markovian providing a suitably optimal choice of reaction coordinate is made, i.e. the reaction coordinate is a good approximation of the committor \cite{krivov_reaction_2013}.

A similar method is Milestoning \cite{faradjian_computing_2004}, which can be used to inform both Markovian and non-Markovian models for the dynamics of the RC from ensembles of trajectories which span  part of the nucleation process. In the context of nucleation this has been combined with MFEP methods to inform the placement of ``milestones'' and used to compute rates of crystallisation from the melt \cite{santiso_general_2015}. However simulations of nucleation in the slow dynamics regime remain relatively unusual compared to studies in the regime where the process is assumed rapid compared to the waiting time between nucleation events.

\section{\label{sec:future}Conclusions and future challenges}

Molecular simulations have been playing an important role in complementing the experimental insight into the kinetics of crystal nucleation for decades. Recent advancements in the field of enhanced sampling, together with the ever-increasing accuracy of the available force fields as well as the unprecedented capabilities of the current high performance computing facilities, all provide exciting new avenues for the computational scientist. At a time when computing crystal nucleation rates is becoming a possibility for many research groups across the globe, we believe it is important to take stock of what we have learned in terms of the limitations of our simulations, particularly in terms of the several aspects that can contribute toward the uncertainty of our results. In this perspective, we focused on atomistic simulations and discussed seven of such aspects in the context of the recent literature.

We started by questioning whether the nucleation events we need to investigate in order to extract the nucleation rate can in fact be considered as ``rare enough'' to justify the assumption of a steady-state nucleation rate. This is key to quantify the uncertainty associated with our estimate of $\mathcal{J}$ via both unbiased and enhanced sampling simulations and we have highlighted a few methodologies there are now available to do just that. Then, we moved onto finite size effects, a long-standing issue that affects atomistic simulations as a whole. These are exciting times for the community, as we are now able to probe the emergence of these spurious effect in models containing up to several million atoms. Related to this aspect is the problem of solute depletion, another practical aspect of molecular simulations that we can now overcome by leveraging a diverse portfolio of computational techniques. 

A slightly less encouraging picture emerged instead from our attempts to accurately model the microscopic kinetics of the system, particularly when it comes to take into account the fact that crystal nucleation is an exothermic process. Only within the last few years the community has started to assess the effect of thermostatting the entire simulation box whilst simulating crystal nucleation - a choice that might play a role in determining whether the nucleation process can be considered as Markovian or not. Indeed, the emergence of non-Markovian RC dynamics is a problematic aspect that clashes with some of the assumptions CNT builds upon - and that we have discussed in this perspective together with some perhaps not entirely obvious limitations of this theory. 

It is impossible not to touch on the choice of the reaction coordinate when it comes to computing crystal nucleation rates. Much has been said in the past, but here we focused on the growing evidence that in many cases we need to acknowledge  the complexity of the nucleation process and construct very specific reaction coordinates starting from a diverse portfolio of order parameters. This is where, we believe, the rise of machine learning can contribute to further the current state of affairs, as a prime tool to identify, select and potentially even learn the most accurate reaction coordinate available to us. We also argue, though, that at this point in time we are running out of excuses for avoiding to validate our reaction coordinates, as we now have both the theoretical tools and - in most cases - the computational power as well.

Finally, we offer a perhaps provocative take on the usage of path-sampling techniques, which are righteously becoming more and more popular as the tool of the trade to compute crystal nucleation rates but that still suffer from a number of limitations. Amongst such limitations is the slow dynamics of the system, which we encounter anytime we work at the strong supercooling or supersaturation regimes we often need in order to make our simulations tractable from the practical standpoint of the computational resources available to us.

At this stage, the reader might think our perspective is painting a rather bleak picture: substantial inconsistencies between experimental and simulated nucleation rate persists; investigating realistic systems at conditions relevant to the experimental reality is, more often than not, still computationally prohibitive; and, even by using state-of-the-art methodologies, our estimates of $\mathcal{J}$ are likely to be affected by a sizable degree of uncertainty, accumulated through a number of diverse aspects, only seven of which we have discussed in some details here. 

However, we believe that the field is a very different, in fact much better spot than it was only a decade ago: this is because the importance of quantifying uncertainty has now been acknowledged across the board, with several research groups going back to the basics and questioning the assumptions we have built our results upon. Of course, this is a often painful process that is bound to uncover some more problematic aspects related to our simulations, but we argue that we should embrace the challenge and work toward promoting the transparency as well as the reproducibility of our work. 

We also feel that the gap between the community working on the so-called model systems and those computational scientists interested instead in more realistic system is quickly narrowing, which in turn should facilitate knowledge exchange in terms of both techniques and expertise. In addition, it is worth pointing out that many of the issues we have addressed in here are by no means exclusive to crystal nucleation. The occurrence of rare events is ubiquitous in the natural sciences~\cite{malik_rare_2020} and quantifying the kinetics characterising these processes is a fundamental open question with reverberations across e.g. climate science~\cite{ragone_computation_2018,webber_practical_2019} as well as epidemiology~\cite{billings_seasonal_2018,grafke_numerical_2019}.

In summary, we hope we have managed with this perspective not only to highlight some problematic aspects, but to showcase some of the excellent work that has recently been done to overcome these challenges and break new ground. The field is moving forward at a very fast pace, leveraging concepts righteously taken from other disciplines - it is our duty to keep an eye on the accuracy of our simulations as we embark in the next chapter of this long-standing endeavour.

\begin{acknowledgments}
We gratefully acknowledge the use of Athena at HPC Midlands+, which was funded by the EPSRC via the grant n.EP/P020232/1, via the HPC Midlands+ consortium, as well as the high-performance computing facilities provided by the Scientific Computing Research Technology Platform at the University of Warwick. We would also like to acknowledge the EPSRC program grant EP/R018820/1, "Crystallisation in the Real World: Delivering Control through Theory and Experiment", as well as the EPSRC Centre for Doctoral Training in Modelling of Heterogeneous Systems (EPSRC grant EP/S022848/1). We gratefully acknowledge Dr. Matteo Salvalaglio for insightful discussions and for reading an earlier version of the manuscript.
\end{acknowledgments}

\section*{\label{sec:app_A}Appendix A}
Details about the LJ calculations reported in Fig. 2.
Simulations were performed in the NPT ensemble using the LAMMPS MD package (version released 5 June 2019). The Lennard-Jones parameter was implemented with a cutoff of $3.5\, \sigma$ and an associated tail correction. A periodic cubic box of 4\,000 liquid atoms was produced by melting a face centered cubic crystal. A small amount of linear quench was then applied to lower the system to the desired temperature before the simulation began. An isotropic pressure of $p\mbox{*}=5.68$ was applied through the simulation using a chain of 5 thermostats attatched to a Hoover barostat of damping parameter $0.5t\mbox{*}$ with an applied Martyna-Tobias-Klein correction. \par
The thermostatted using a chain of 5 Nos\'e-Hoover thermostats of damping coefficient $0.1t\mbox{*}$. The size of the cluster was calculated using the ten Wolde $q_6$ order parameter\cite{ten_wolde_numerical_1995} as implemented in the molecular dynamics package LAMMPS (we are aware of the issue with $q_6(n)$ in LAMMPS, but errors from this source should be small by the time the critical size is reached. 
The system was observed over 5\,000\,000 timesteps of length $t\mbox{*}=0.002$, with $q_6(n)$ being recorded every 100 timesteps. \par
Simulations (258 for $T\mbox{*}=0.68$; 224 for $T\mbox{*}=0.765$; and 823 for $T\mbox{*}=0.8$) were analysed in python, by fitting a sigmoid curve to the value of $q_6(n)$. If this never exceeded 2\,000 the simulation was determined not to have crystallised. Shifted curves were created by removing the time associated with the midpoint of the sigmoid curve from the timesteps of the simulations. \par
The survival probabilities were found by dividing the output of simulations into 5 approximately equal-sized sets. The location of the midpoints of the sigmoid curves were binned into 15 bins ranging from $t\mbox{*}=0$ to the maximum value of $t\mbox{*}$ at which a midpoint was recorded. The plotted values are the mean of the values for these 5 sets, with the error bar corresponding to the error in the mean (standard deviation divided by the sqare root of the number of sets). Exponentials of the form $\exp\left[-\left(\mathcal{J}t\mbox{*}\right)^\alpha\right]$, with $\alpha$ either allowed to vary or constrained to 1, were fitted using {\tt scipy.optimize.curve\_fit}. Bounds were imposed to ensure a reasonable fit was given, with details of these bounds available in the analysis code. For $T\mbox{*}=0.68$, all errors on points were 0, so errors were neglected. For $T\mbox{*}=0.8$ only the error on the initial point was 0, and for $T\mbox{*}=0.765$ errors on the first and last point were 0; in these cases fitting was performed on data points with associated errors, and data points with 0 error were omitted. \par 
Analysis and input scripts can be found on github.

\section*{\label{das}Data availability statement}
The data we have used to put together Fig.~\ref{FIG_5} are available from the corresponding author upon reasonable request.

\section*{\label{credit} Note} 
The following article has been submitted to the Journal of Chemical Physics. After it is published, it will be found at https://aip.scitation.org/journal/jcp.

\bibliography{ms}

\end{document}